**Correctness Is Not Homogeneous Evidence: A Correctness-conditioned Evidence-aware Knowledge Tracing Model**

**Abstract:** Knowledge tracing models typically use response correctness as a central observation for estimating students' latent knowledge states. However, the same correct or incorrect response may arise from different behavioral contexts, such as rapid guessing, hint use, repeated attempts, or different response processes. Treating correctness as uniformly informative may therefore introduce ambiguity into recurrent state updates. To address this issue, this study proposes a Correctness-conditioned Evidence-aware Knowledge Tracing model (CE-KT). The model first derives weakly supervised behavioral proxy scores from observable response-process features, including response time, hint use, attempt count, and behavioral history. These scores are used as conditional behavioral signals rather than direct measures of mastery, response quality, or cognitive state. CE-KT then separates correct and incorrect interactions through two correctness-specific gates. Each gate performs full-state modulation on both the LSTM hidden state and cell state, and the modulated states are fed back into subsequent recurrent updates. The model was evaluated on ASSISTments data through four research questions concerning behavioral heterogeneity, prediction bias, recurrent state modulation, and mechanism-level differentiation. The results showed that behavioral condition scores were associated with future same-skill performance within fixed correctness groups. The association was clearer for incorrect interactions. For correct interactions, the difference was small but negative, which indicates behavioral heterogeneity but does not support interpreting higher scores as stronger evidence of stable mastery. CE-KT generally outperformed behavior-fusion alternatives on the main predictive metrics, although the calibration advantage was not consistently supported. Additional ablations and mechanism analyses provided partial support for correctness-specific recurrent modulation and recurrent feedback. These findings suggest that behavioral information can be useful for conditioning the interpretation of correctness during recurrent state updates. However, the proposed proxy scores should not be interpreted as direct measures of mastery or productive error, and the observational results do not establish causal effects on learning. The correct-side proxy, in particular, requires reconstruction and further validation across datasets and analytic settings.

## 1. Introduction

Knowledge tracing (KT) is a core modeling task in educational data mining. It uses students' historical response interactions to estimate their latent knowledge states and to predict their future response correctness (Corbett & Anderson, 1994). The growth of online learning platforms has made large-scale interaction logs increasingly available, including response correctness, response time, hint use, and attempt count (Zhao & Yin, 2021). By dynamically modeling students' knowledge states, KT provides a technical basis for intelligent tutoring systems, adaptive learning, personalized resource recommendation, learning-path planning, and performance prediction (Adhikary et al., 2026).

Despite the development of recurrent, memory-augmented, and attention-based KT models, binary response correctness remains a central observation for knowledge-state modeling. However, correctness is not an error-free signal of latent knowledge. The same correct response may arise from independent knowledge, rapid guessing, or hint-assisted problem solving, whereas the same incorrect response may reflect a knowledge gap or a careless slip (Corbett & Anderson, 1994; San Pedro et al., 2011). Contextual extensions of Bayesian knowledge tracing further show that the probabilities of guessing and slipping may vary across response contexts (Baker et al., 2008). We refer to this phenomenon as correctness observation heterogeneity: the same binary correctness label may carry different evidential value for subsequent state updating depending on the response process. Here, evidential value denotes the relative information that a correctness observation provides for model state updating under a behavioral context; it is not an objective measure of the learner's true cognitive state.

Prior work has addressed this heterogeneity from three main directions. First, contextual guess-and-slip models adjust observation probabilities according to response context, showing that the meaning of correct and incorrect responses is not fixed (Baker et al., 2008; Wang & Heffernan, 2012). Second, behavior-enhanced deep KT models encode response time, hints, attempts, or other process features into the input representation to enrich hidden states (Zhang et al., 2017; Shin et al., 2021). Third, state-observation separation approaches distinguish the evolution of knowledge states from the generation of visible responses (Loong & Chang, 2024). These studies explain how observation probabilities may vary, whether behavioral information can be useful, and why state and observation should be separated. They do not yet specify how current behavioral context should modulate the way a correctness observation is written into a recurrent state, or how this modulation should continue to affect subsequent states.

The central limitation is therefore not simply that existing KT models ignore behavioral information. The unresolved issue is the modeling role assigned to that information. Behavioral features may describe aspects of a learner's history, but they may also condition the evidential value of the current correctness observation. When behavior, skill, and correctness are merged as ordinary inputs, these two roles are mixed within the same hidden representation. When behavior is used only at the output layer, its influence does not become part of the recurrent history. Moreover, applying the same modulation rule to correct and incorrect responses fails to reflect their different update directions. A clearer mechanism is needed to let behavioral features condition how correct and incorrect observations affect state updates and how the resulting modulation is propagated through later interactions.

To address this gap, this study treats response-process behavior as conditional information for interpreting current response correctness and proposes a Correctness-conditioned Evidence-aware Knowledge Tracing model (CE-KT). The model first derives behavioral proxy scores from response time, hint use, attempt count, and related process features. These proxy scores are weakly supervised behavioral signals; they are not direct measures of mastery, response quality, or cognitive state. CE-KT then uses current response correctness to select a correct-response or incorrect-response gate. The selected gate performs full-state modulation by adjusting both the LSTM hidden state and cell state, thereby controlling how strongly the current interaction is written into the recurrent state. The modulated state is then fed back to the subsequent sequence, allowing the behavioral interpretation of the current observation to persist beyond the current prediction.

This mechanism leads to four research questions:

1. RQ1: After separating interactions by current response correctness, are behavioral condition scores within correct and incorrect interactions associated with future same-skill performance?
2. RQ2: Do standard KT models with different sequence-modeling mechanisms show systematic prediction bias across strata of behavioral proxy scores?
3. RQ3: Does correctness-specific recurrent full-state modulation outperform alternative behavior-fusion strategies, and to what extent does its performance depend on recurrent feedback, gate structure, and gate-input information?
4. RQ4: Can the model treat interactions with the same current correctness differently according to their behavioral context?

## 2. Related Work

### 2.1 Correctness Observation Heterogeneity and Interaction-Level Evidential Differences

KT infers an unobservable latent knowledge state from visible response correctness (Corbett & Anderson, 1994). Response correctness is an observation generated under a particular item and response condition; it should not be equated with the latent knowledge state itself (Loong & Chang, 2024). Bayesian knowledge tracing and its contextual extensions make this distinction explicit through guess and slip parameters, showing that correct and incorrect observations can vary in how strongly they inform judgments about mastery (Corbett & Anderson, 1994; Baker et al., 2008). On this basis, this study uses interaction-level evidential differences to refer to relative differences in the evidential value of correctness after holding current response correctness fixed. This concept is grounded in observational modeling with behavioral proxies; it is not a direct measurement of a learner's true cognitive or psychological state.

Behavioral context can provide process information that is absent from a binary correctness label. Very short response times may be associated with rapid guessing or low response effort, and correct responses after hints or repeated attempts may provide weaker evidence of independent solution behavior (Wise & Kong, 2005; Baker et al., 2004). These interpretations, however, are context dependent. Response time, hint use, and attempt count are influenced by learner ability, item difficulty, task design, and platform logging rules, so they cannot be directly interpreted as mastery, response quality, or cognitive effort (Kong et al., 2007; Pelanek, 2024). Similarly, although incorrect responses are often treated as negative observations, errors under appropriate feedback and processing conditions can still be associated with later understanding (Kapur, 2014; Metcalfe, 2017). Therefore, the evidential value of correctness in this study denotes only a relative modeling construct under fixed correctness, not a psychological diagnosis.

## 2.2 Behavioral Information in KT: Modeling Roles, Functions, and Limitations

Existing KT research has incorporated behavioral information in three broad ways. The first line adjusts the observation model. Contextual guess-and-slip approaches estimate the probabilities of correct guesses and incorrect slips from the response context (Baker et al., 2008), and response-time extensions of BKT use first response time to improve the explanation and prediction of current and future performance (Wang & Heffernan, 2012). The second line treats behavior as additional model input. Deep KT models can encode item properties, response time, time intervals, attempts, or other process features into the input or response representation (Zhang et al., 2017; Shin et al., 2021; Sun et al., 2022). The third line separates state dynamics from observable response generation, modeling how knowledge states evolve and how those states are mapped to visible outcomes (Loong & Chang, 2024).

These approaches establish the usefulness of behavioral features, but they leave a functional ambiguity. Once behavior, skill, and correctness are jointly encoded into a hidden representation, improved prediction does not reveal whether behavior enriches the model's summary of latent knowledge or instead conditions the meaning of the current correctness observation. If behavior is introduced only at the output layer, its effect is limited to the current prediction and does not become part of the recurrent history. In addition, response time, hint use, and attempt count have context-dependent meanings and cannot be read as direct indicators of mastery or response quality (Pelanek, 2024). A structurally testable mechanism is therefore needed to specify what behavioral information explains, where it acts, and how its effect is propagated.

This gap can be expressed as three linked questions. First, does behavioral information describe the learner's latent knowledge state, or does it condition the evidential value of the current correctness observation? Second, where should behavioral information act: during state updating, during observation mapping, or only at the prediction output? Third, does behavioral information affect only the current prediction, or does it change how the current interaction is carried into subsequent recurrent states?

## 2.3 Potential of Evidence-Aware Recurrent Gating

The proposed mechanism draws on three related ideas: state-observation separation, conditioned gating, and recurrent state propagation. State-observation separation provides the theoretical basis for assigning behavioral information a specific modeling role. In statistical state-space models, latent-state dynamics are distinguished from the formation of visible observations, which helps separate state change from observation uncertainty (Durbin & Koopman, 2012). In KT, this perspective appears in the original distinction between latent knowledge and visible responses in BKT (Corbett & Anderson, 1994), in contextual guess-and-slip models where observation meaning varies by context (Baker et al., 2008), and in CtrKT, which separately models knowledge-state change and response-score generation (Loong & Chang, 2024). This line of work supports treating behavioral information as a condition for interpreting the evidential value of correctness rather than as a direct measure of knowledge.

Conditioned gating provides a technical basis for modulating state updates with behavioral context. LSTM showed that recurrent models can use gates to selectively control information writing, retention, and output (Hochreiter & Schmidhuber, 1997). The forget gate further allows this control to depend dynamically on the current input and previous state (Gers et al., 2000). FiLM extends conditional modulation to intermediate representations by using external conditions to generate feature-wise scaling and shifting parameters (Perez et al., 2018). Together, these mechanisms show that external context need not be appended only as an ordinary input feature; it can also be transformed into modulation parameters that selectively change how current information enters and remains in the model's internal state.

Recurrent state propagation explains why the location of modulation matters. DKT uses a recurrent hidden state to integrate prior interactions and pass the updated state to later time steps, which means that an interaction can continue to influence future predictions only if its information enters the recurrent state (Piech et al., 2015). LSTM and its gating mechanisms provide a concrete way to retain, update, or remove historical information within this recurrent state (Hochreiter & Schmidhuber, 1997; Gers et al., 2000). If behavioral information modulates only the current output, its effect stops at the current prediction. By contrast, feeding the modulated hidden state and cell state into the next time step allows the behavioral interpretation of the current correctness observation to influence subsequent state updating.

## 3. Method

This study proposes Correctness-conditioned Evidence-aware Knowledge Tracing (CE-KT). The model first derives conditional behavioral signals for the current response correctness label from observable response-process features. It then uses the current response result to select a correct-response or incorrect-response gate, modulates both the LSTMCell hidden state and cell state, and feeds the modulated states to the next time step. The term evidence-aware means that the model uses behavioral context to distinguish interactions with the same correctness label; it does not claim to measure objective response quality.

### 3.1 Problem Definition and Overall Framework

#### 3.1.1 The KT Task and the Additional Problem Addressed

Let the t-th interaction of student i be denoted by $(s_{it}, y_{it}, X_{it})$. Here, $s_{it}$ is the skill or knowledge component practiced in the current interaction, $y_{it} \in \{0,1\}$ is the current response correctness label, and $X_{it}$ denotes response-process behavior observable by the end of the current problem log, such as response time, hint use, and attempt count. $X_{it}$ does not include information after the current interaction.

$$\mathcal{H}_{it} = \{(s_{i\tau}, y_{i\tau}, X_{i\tau})\}_{\tau \le t},$$

The standard KT objective is to predict the probability that the student will answer the next target interaction correctly:

$$\hat{y}_{i,t+1} = P(y_{i,t+1} = 1 \mid \mathcal{H}_{it})$$

In DKT, skill and correctness are usually encoded as the current interaction input, and a recurrent state summarizes the history:

$$x_{it} = \text{Embed}(s_{it}, y_{it}),$$
$$h_{it} = \text{LSTM}(x_{it}, h_{i,t-1}), \qquad \hat{y}_{i,t+1} = \sigma(W_{s_{i,t+1}} h_{it} + b_{s_{i,t+1}})$$

In this study, the hidden state $h_t$ is treated as the model's internal summary of the response history, not as a direct measurement of the learner's true knowledge level. The additional question is how the current correctness observation $y_{it}$ should be incorporated into later recurrent states once the response result is already observed. The question is not whether the learner truly mastered the skill, but whether response-process information can condition the evidential value of the current correctness observation for subsequent state updating.

#### 3.1.2 Core Mechanism and Functional Positioning

CE-KT assigns behavioral information a restricted function: it conditions the evidential value of correctness. This value indicates the relative information that an interaction provides for later model-state updating after current correctness is fixed. It is not a direct measure of knowledge, effort, guessing, slipping, response quality, or psychological state.

The mechanism has three steps. First, behavioral proxy scores provide the conditioning signal. Second, current response correctness selects separate correct-response and incorrect-response gate paths. Third, the post-modulation hidden state and cell state are carried forward through recurrent feedback. Therefore, behavioral information is not merged indiscriminately with skill and correctness as an ordinary LSTM input. Instead, it enters as a correctness-specific observation gate that modulates how the current interaction is written into the recurrent state.

3.1.3 Model Structure

Figure 1. Overall CE-KT architecture and single-step information flow

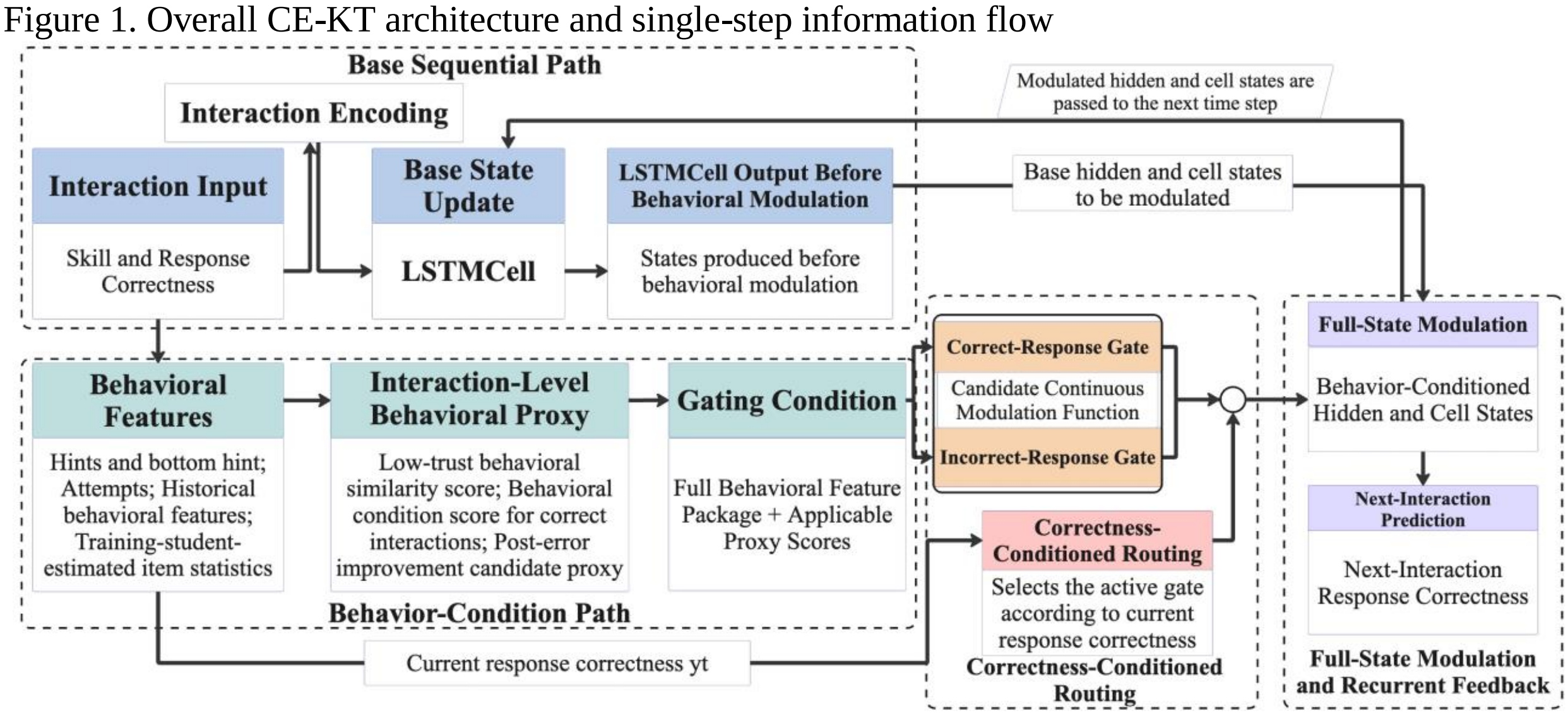

The model contains a base sequential path, a behavioral conditioning path, correctness-conditioned routing, and recurrent full-state modulation. The base path uses skill, correctness, and the previous recurrent state to produce pre-modulation states. The behavioral path uses current process features and causal historical summaries to form a gate-conditioning vector. Current correctness selects the correct-response or incorrect-response gate, and the selected gate modulates the pre-modulation states. The resulting states are used for next-interaction prediction and are also fed back to the next time step. State refers specifically to the LSTMCell hidden state and cell state unless otherwise stated.

3.2 Behavioral Feature Construction and Temporal Information Use

3.2.1 Raw Behavioral Variables

This study uses response time, hint count, bottom-hint flag, and attempt count from the ASSISTments 2012-2013 problem logs. Each retained problem log is treated as one response interaction, lower-level actions are not re-aggregated. The response correctness label is taken from the same-row correct field, and ms_first_response, hint_count, bottom_hint, and attempt_count are also taken from the corresponding problem-level summary fields.

A critical logging limitation is that the problem-level file does not fully preserve the within-problem action order. Therefore, correct = 1 cannot be interpreted as first-attempt correctness, and bottom_hint = 1 does not prove that the learner viewed the bottom hint before producing a correct response. These variables are used only as behavioral proxies associated with the response result. No causal or temporal claim is made about the order of hint use and correctness within a problem log. This study treats these variables solely as behavioral proxies associated with response outcomes, without drawing further inferences regarding their temporal order or causal relationships.

3.2.2 Temporal and Contextual Processing of Behavioral Features

Raw response times in learning environments are typically right-skewed, and log transformation is commonly used (Pelanek, 2024). The training-set distributions in this study also showed strong right skew, so response time was transformed as log(1 + RT) to reduce the influence of extreme long responses.

Because students differ in their baseline response speed, log response time was standardized within student using only the causal prefix before the current interaction. The resulting value represents whether the current response is faster or slower than the student's own prior response pattern, not whether the student is fast or slow relative to all students. When a student had insufficient history, fallback means and standard deviations estimated from the training set were used. If the student's historical response-time standard deviation was close to zero, the historical mean was retained and the training-set standard deviation was used for scaling.

For hints and attempts, the model uses item-level relative features by comparing the current value with the mean for the same problem estimated from training students. These item statistics are estimated only from the training set and are frozen for validation and test students. This processing helps distinguish items that normally require more hints or attempts from behavior that is unusual for a particular interaction.

3.2.3 Separation of Training, Runtime, and Validation Information

To avoid future information leakage, CE-KT separates three information windows: runtime model input, the training-student window used to construct weak-supervision anchors, and the non-overlapping future window used to examine whether the proxies are associated with later same-skill performance. At time t, runtime input contains only information available by the end of the current problem log. Student-level behavioral summaries use only the causal prefix before t, and item statistics are estimated from training students and then frozen.

For training students, the first to third subsequent same-skill opportunities are used only to define weak-supervision anchors. The fourth to eighth subsequent same-skill opportunities are reserved for non-overlapping future-window validation. This separation avoids reusing the same future observations for both proxy construction and validation. It does not turn the proxy into a direct measure of cognitive state, and it does not support causal interpretation.

Figure 2. Temporal relationship among online model input, weak-label windows, and independent-validity windows

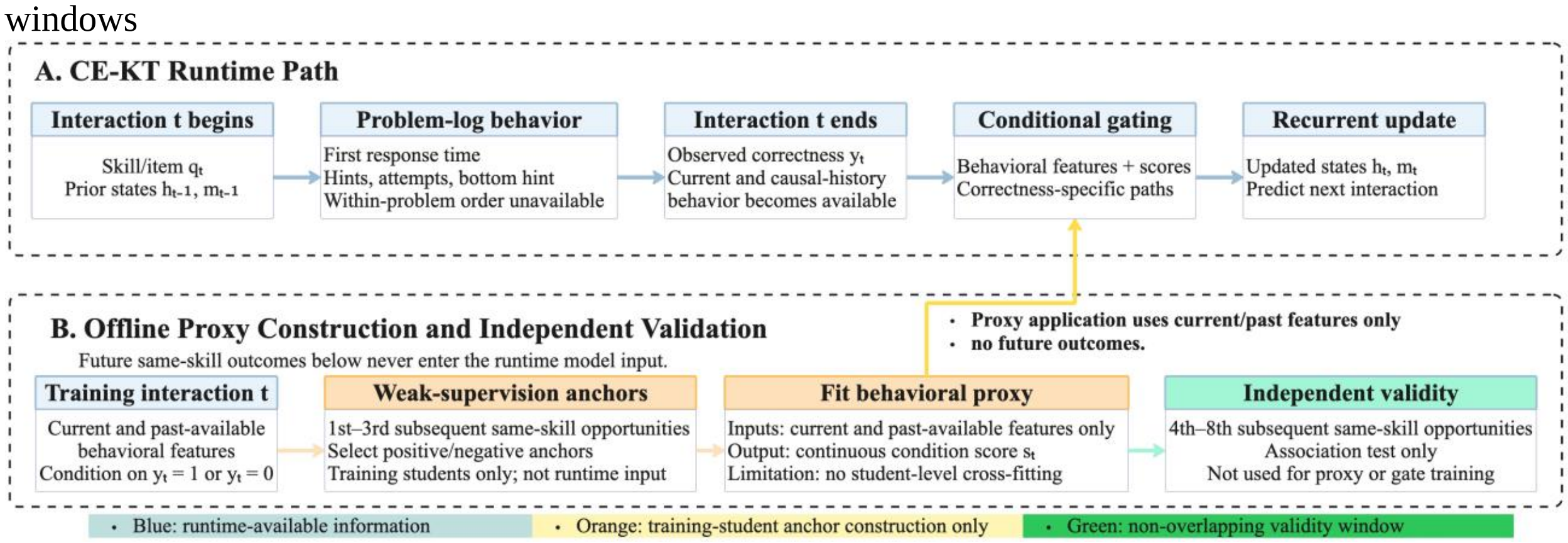

3.3 Behavioral Proxy Construction and Gate-Conditioning Vectors

To avoid naming the signals in a way that presupposes psychological meaning, this study refers to the three gate-conditioning signals as behavioral condition scores: the low-trust behavioral similarity score, the positive-interaction behavioral condition score, and the negative-interaction behavioral condition score. The first summarizes similarity to heuristic low-trust rules. The latter two summarize behavioral-context differences within currently correct and currently incorrect interactions. All three are weakly supervised behavioral proxy scores, not direct measures of mastery, response quality, guessing, or productive learning.

Because true labels are unavailable, weak-supervision anchors are selected using heuristic rules. Positive anchors are typical examples of weak label 1, and negative anchors are contrastive examples of weak label 0. They do not correspond to current correctness, incorrectness, or true cognitive states. Intermediate samples do not participate in proxy training, but they still receive continuous scores at application time. Future same-skill performance is used only for anchor selection among training students and does not enter runtime model input. The gate retains both the full behavioral feature vector and the three behavioral condition scores.

3.3.1 Low-Trust Behavioral Similarity Score

The low-trust behavioral similarity score identifies behavioral patterns whose correctness observations are expected to have weaker evidential value. It does not diagnose true guessing, slipping, or non-cognitive state. Its positive anchors are formed from rapid correct responses, co-occurrence of correctness and bottom-hint flags within the same problem log, and anomalously high attempts followed by correctness. Interactions not matched by these rules are not automatically treated as trustworthy; correct and incorrect interactions can both enter the contrastive candidate pool. Therefore, this score mainly learns similarity to low-trust correct-response rules and does not establish a separately validated low-trust error category.

For formula compactness, $a_t$ denotes the current attempt count, $H_t^{\text{bottom}}$ denotes the bottom-hint flag, and $\bar{a}_q$ denotes the mean attempt count for the same problem estimated from training students. The three rule indicators are denoted as $q_{\text{quick}}$, $q_{\text{bottom}}$, and $q_{\text{attempt}}$.

$$q_t^Q = 1\{RT_t \leq Q_{0.05}(RT_{i,<t})\} \cdot 1\{a_t = 1\} \cdot 1\{y_t = 1\}$$

$Q_{0.05}(RT_{i,<t})$ is the fifth percentile of the student's prior response-time distribution before time $t$. It identifies correct responses in the student's own extreme low tail rather than using a fixed number of seconds. The rule also requires one attempt and a correct result, thereby excluding extremely fast incorrect responses and cases that became correct only after repeated attempts. The fifth percentile is an operational weak-supervision threshold, not a psychological diagnosis of guessing. In sensitivity analyses, changing this percentile to 1%, 5%, and 10% and retraining the model did not produce stable differences in the main KT metrics, and the direction of the negative-side validation result remained consistent.

$$q_t^H = 1\{u_t \geq 1\} \cdot 1\{y_t = 1\}$$

This rule captures co-occurrence of a correct result and a bottom-hint flag in the same problem log. Because the within-problem action order is not reconstructed, the text should not claim that the student answered correctly after viewing the bottom hint. Hint use may also serve a learning function. The rule only supplies low-trust positive-anchor candidates and does not treat all bottom-hint use as gaming. This rule matched 82 interactions, approximately 0.018% of the modeled interactions, and therefore contributed little by count.

$$q_t^A = 1\left\{\frac{a_t}{\mu_{a,j}^{\text{tr}}} > 3\right\} \cdot 1\{y_t = 1\}$$

The attempt rule uses the same-problem mean attempt count estimated from training students and frozen for validation and test students. Raw attempt counts are affected by item design and difficulty, so the rule uses relative attempts rather than an absolute attempt count. The 90th, 95th, 97.5th, and 99th percentiles of relative attempts were 1.52, 1.94, 2.55, and 3.39, respectively. Thus, the threshold of 3 lies in the extreme high tail of the empirical distribution and is a conservative screening rule for typical low-trust positive anchors, not a universal anomaly boundary. In the analytic sample, this rule did not match any interaction and was retained only as a prespecified safeguard.

The construction of low-trust weak labels proceeds through rule aggregation, contrast-sample selection, and probabilistic modeling. The three rules are combined by logical OR: an interaction is assigned to the positive-anchor set if any rule is triggered. Interactions that do not trigger these rules are not automatically treated as trustworthy. Instead, contrast samples are drawn from interactions with relatively neutral affective conditions. The concentrating, bored, confused, and frustrated variables come from model-predicted probabilities provided in the ASSISTments 2012-2013 School Data with Affect release; they are not self-reports, human observations, or affect labels collected in this study. These affect-related fields are used only to improve the typicality of weak-supervision anchors and are not interpreted as students' true emotional states. After rule aggregation and contrast-sample selection, a standardized logistic regression maps observable behavior to a continuous rule-similarity score.

Rule-matched interactions remain in the original sequence to avoid changing the actual response history and to allow the model to learn how to handle similar interactions at runtime. In the analytic sample, the three rules matched 23,843 interactions, accounting for 5.34% of retained interactions. Rapid-guessing matches accounted for 23,761 of these cases and were therefore the dominant source. Accordingly, $g_t$ should be interpreted mainly as similarity to rapid-guessing-type low-trust rules rather than as a general measure of all non-cognitive behavior. The internal AUC of the rule-similarity model was 0.9926, indicating that logistic regression reproduced the heuristic labels defined by overlapping behavioral features well; this internal result does not establish independent validity for detecting true non-cognitive behavior.

3.3.2 Positive-Interaction Behavioral Condition Score

The positive-interaction behavioral condition score is defined only when the current response is correct. It summarizes behavioral-context differences among correct interactions. Weak-label anchors combine the first to third subsequent same-skill opportunities with current and historical process features, but the score is applied using only current and past available features. Having at least two subsequent opportunities provides only minimal repeated observation and is not evidence that the learner has mastered the skill.

To summarize performance on neighboring same-skill responses before and after the current interaction, this study assigns distance-decay weights according to their distance from the current interaction. Same-skill responses closer to the current interaction receive larger weights, whereas more distant responses receive smaller weights. For up to three same-skill responses before and after the current interaction, the weight of the $d$-th neighboring opportunity is defined as:

$$w_d = 0.6^{d-1}, d = 1,2,3.$$

Let $y_{t,d}^-$ and $y_{t,d}^+$ denote the response correctness outcomes of the $d$-th same-skill opportunity before and after the current interaction, respectively. The local weighted performance before and after the current interaction is defined as:

$$P_t^- = \frac{\sum_{d=1}^{m_t^-} w_d \, y_{t,d}^-}{\sum_{d=1}^{m_t^-} w_d}.$$

$$P_t^+ = \frac{\sum_{d=1}^{m_t^+} w_d \, y_{t,d}^+}{\sum_{d=1}^{m_t^+} w_d}.$$

Here, $d$ denotes the ordinal distance from the current interaction among same-skill opportunities, not the number of elapsed days. $m_t^-$ and $m_t^+$ denote the numbers of same-skill opportunities actually included before and after the current interaction, respectively; both have a maximum value of 3. Opportunities closer to the current interaction receive larger weights. The local performance change is defined as:

$$\Delta_t^{(s)} = P_t^+ - P_t^-$$

The positive anchor $M_t = 1$ for the positive-interaction behavioral condition score is defined only for $y_t = 1$, with at least two subsequent same-skill opportunities, and requires:

$$M_t = 1 \mid\Leftarrow \quad P_t^+ \geq 0.70, \mid \Delta_t^{(s)} \geq -0.05, \mid g_t < 0.35, \mid u_t = 0$$

Positive-anchor selection uses two layers: subsequent performance and response process. A subsequent weighted correctness of at least 0.70 and a local change of at least -0.05 retain interactions with relatively high subsequent performance and no marked decline. The conditions $g_t < 0.35$ and bottom-hint = 0 reduce the influence of low-trust behavior and answer-type hints. Thus, the positive anchor requires both relatively stable subsequent performance and no clear low-trust feature in the current process. These thresholds are operational settings for weak-supervision sample selection, not mastery-diagnostic criteria. For the contrast class, the negative anchor uses the same candidate scope and satisfies at least one of the following conditions:

$$M_t = 0 \mid\Leftarrow \quad (P_t^+ \leq 0.50) \vee (\Delta_t^{(s)} \leq -0.20) \vee (g_t \geq 0.65) \vee (u_t \geq 1)$$

The negative anchor $M_t = 0$ identifies correct interactions with low subsequent performance, a marked local decline, high similarity to low-trust behavior, or bottom-hint use. Because any one of these conditions may indicate a correct response with limited evidential value, the negative anchor is constructed by logical OR. Intermediate samples that satisfy neither the positive nor the negative criteria are excluded from weak-label training but still receive continuous scores when the proxy is applied. After class balancing, a second-order polynomial logistic regression is trained using current and past available behavioral features, item statistics, and same-skill history. Its output is defined as the probability that, conditional on the current response already being correct, the interaction resembles the positive anchor:

$$r_t^p os = P(M_t = 1 | X_t, y_t = 1)$$
$$c_t^p os = r_t^p os(1 - g_t), y_t = 1$$

The factor $1 - g_t$ downweights the positive-anchor similarity probability $r_t^{pos}(1 - g_t)$ according to the degree of low-trust behavior. Compared with setting low-trust interactions to zero under a fixed threshold, this retains the proxy model's probabilistic assessment of low-trust intensity. The positive-interaction proxy has an internal AUC of 0.7034, indicating that it can reproduce the behavioral pattern expressed by the weak labels to some extent. This internal AUC reflects reproduction of the weak-supervision anchors only and does not establish independent validity. Whether the positive-interaction behavioral condition score is associated with later same-skill performance not used to construct the weak labels is tested in RQ1 using the non-overlapping fourth-to-eighth subsequent same-skill opportunities.

### 3.3.3 Negative-Interaction Behavioral Condition Score

The negative-interaction behavioral condition score is defined only when the current response is incorrect. It estimates how similar the current error is, in behavioral-feature space, to a post-error improvement candidate. This term refers only to an interaction that is currently incorrect and shows improved same-skill performance after the error within the observation window. It is an operational name for weak-supervision anchor selection; it does not mean that the error caused subsequent improvement and is not equivalent to a true productive error. The most recent three same-skill opportunities are used to balance local-performance stability against confounding from an excessively long time span. The decay coefficient 0.6 is an operational parameter rather than a theoretical constant.

Candidate interactions are limited to training students with a current incorrect response ($y_t = 0$), no low-trust rule match, and at least two subsequent same-skill opportunities. The positive anchor for the negative-interaction behavioral condition score requires:

$$PE_t = 1 \mid\Leftarrow \quad \Delta_t^{(s)} \geq 0.20, \mid P_t^{+} \geq 0.60, \mid g_t < 0.30$$

The negative-side positive anchor combines subsequent performance with the response process. A local change of at least 0.20 and a post-error same-skill weighted correctness of at least 0.60 select incorrect interactions with a relatively clear improvement that reaches a minimum level. The condition $g_t < 0.30$reduces interference from low-trust behavior. The negative anchor is restricted to currently incorrect responses with no local improvement or with future performance no higher than past performance, while requiring $g_t < 0.50$:

$$PE_t = 0 \mid\Leftarrow \quad \left\{(\Delta_t^{(s)} \leq 0) \vee (P_t^{+} \leq P_t^{-})\right\} \wedge (g_t < 0.50)$$

Intermediate samples that satisfy neither the positive nor the negative criteria are excluded from weak-label training but still receive continuous scores when the proxy is applied. After class balancing, a logistic regression with second-order terms is trained using current and past available behavioral features, and its output is defined as the raw negative-side probability $r_t^{neg}$:

$$r_t^n eg = P(PE_t = 1 | X_t, y_t = 0)$$
$$c_t^{neg} = r_t^{neg}(1 - g_t), \quad y_t = 0$$

Here, $r_t^{neg}$ represents the weak-label probability for post-error improvement candidacy. The model input includes the low-trust behavior similarity score, response-time deviation, hints and attempts, bottom-hint use, and behavioral-history summaries. The internal AUC of the negative-interaction proxy is 0.6236. Independent-validity testing uses non-overlapping fourth-to-eighth subsequent same-skill opportunities;

the student-bootstrap high-minus-low group difference is 0.0913, with a 95% confidence interval of [0.0101, 0.2134] and $p < 0.0033$. This supports an association between the score and differences in later post-error performance, but it does not support a causal interpretation.

### 3.3.4 Gate-Conditioning Vector

The low-trust behavior similarity score, the positive-interaction behavioral condition score, and the negative-interaction behavioral condition score do not replace the original behavioral information. Instead, they are combined with the complete behavioral feature vector to form the gate-conditioning vector. Let $\mathrm{b}_t$ denote the complete behavioral vector for the current interaction, including student-normalized response-time deviation, hint use, relative attempt count, bottom-hint use, item statistics estimated from training students, and causal historical features. Let $\mathrm{z}_t$ denote the final gate-conditioning vector:

$$\mathrm{z}_t = [\mathrm{b}_t; c_t^{pos}; c_t^{neg}; 1 - g_t]$$

The vector $\mathrm{b}_t$ retains behavioral evidence that has not been fully compressed by the proxy models. $c_t^{pos}$ and $c_t^{neg}$ provide combined positive- and negative-side behavioral proxy scores, and $1 - g_t$ represents the relative degree to which the current interaction differs from low-trust rule patterns. When the current response is incorrect, $c_t^{pos}$ is set to 0; when the current response is correct, $c_t^{neg}$ is set to 0, so that an inapplicable proxy does not participate in the current gate. This operation is a structural mask only and has no behavioral or knowledge-state meaning. Retaining the complete behavioral package reduces the gate's dependence on weak-supervision proxy error, while retaining the three behavioral condition scores establishes an explicit connection between behavioral information and the observation-interpretation concept. In this section $c_t^{pos}$ and $c_t^{neg}$ always denote behavioral proxies; the later LSTM memory state is marked with complete subscripts or the term memory state to avoid notational confusion.

## 3.4 Correctness-Specific Conditioned Gating

### 3.4.1 Pre-Modulation Recurrent State

The final model keeps skill and response correctness as the basic DKT sequence input. Their joint encoding forms the current-interaction embedding $x_t$. At time $t$, LSTMCell first reads the hidden state and cell state that were modulated at the previous time step and generates the current raw states before behavioral modulation:

$$x_t = Embed(s_t, y_t).$$
$$(h_t^{raw}, cell_t^{raw}) = \mathrm{LSTMCell}(x_t, h'_{t-1}, cell'_{t-1}).$$

This step preserves the base sequential backbone's ability to learn historical dependencies from skills and correctness. The gate-conditioning vector does not replace LSTMCell state updating; it modulates the states after the raw update has been generated.

### 3.4.2 Correct-Response and Incorrect-Response Gates

The correct-response gate and incorrect-response gate read the same conditioning vector but use independent parameters to generate modulation parameters:

$$[\gamma_t^p os, \beta_t^p os] = MLP_p os(z_t).$$
$$[\gamma_t^n eg, \beta_t^n eg] = MLP_n eg(z_t).$$

Current response correctness acts as a discrete routing variable:

$$\gamma_t = y_t \gamma_t^{pos} + (1 - y_t)\gamma_t^{neg}$$
$$\beta_t = y_t \beta_t^{pos} + (1 - y_t)\beta_t^{neg}.$$

When $y_t = 1$, the correct-response gate is used; when $y_t = 0$, the incorrect-response gate is used. The purpose of two gates is not to impose that correct responses must be strengthened or incorrect responses must be weakened. It allows the model to learn different conditioned modulation functions for the two observation types.

3.4.3 Scale-and-Shift Modulation

The gate networks generate feature-wise scaling and shifting parameters for the LSTMCell pre-modulation hidden state and cell state. Scaling controls the strength of each state dimension, whereas shifting adds a bounded context-dependent offset:

$$h'_t = h_t^r aw \odot [1 + tanh(\gamma_t^h)] + tanh(\beta_t^h).$$
$$cell'_t = cell_t^r aw \odot [1 + tanh(\gamma_t^c)] + tanh(\beta_t^c).$$

Here, $\odot$ denotes element-wise multiplication. $\gamma_t^h$ and $\beta_t^h$ modulate the hidden state, whereas $\gamma_t^c$ and $\beta_t^c$ modulate the cell state. Because tanh has range [-1, 1], when a = 0.5, the multiplicative scaling factor is restricted to approximately 0.5 to 1.5. This allows the gate to moderately adjust the contribution of different dimensions in the pre-modulation hidden and cell states while avoiding complete replacement or excessive amplification of the base state.

3.4.4 Post-Modulation State and Prediction

The post-modulation hidden state is passed through a skill-specific sigmoid output layer for the next target skill to generate the next-response correctness probability:

$$\hat{y}_{t+1} = \sigma(W_{s_{t+1}} h'_t + b_{s_{t+1}}).$$

At each time step, the model first uses current skill, current correctness, and the previous recurrent state to form pre-modulation hidden and cell states. It then selects the correct-side or incorrect-side gate according to current correctness and uses the gate-conditioning vector to modulate the pre-modulation states. Behavioral information is therefore not directly concatenated with skill and correctness as the base LSTMCell input; it conditions how the current interaction is written into the recurrent state.

3.5 Recurrent Feedback and Complete Update

The key difference between CE-KT and output-only modulation is recurrent feedback. The post-modulation states $h'_t$ and $cell'_t$ are passed to the next time step:

$$(h_{t+1}^r aw, cell_{t+1}^r aw) = LSTMCell(x_{t+1}, h'_t, cell'_t).$$

If the gate changed only the hidden representation before the prediction head, the next time step would still receive an unmodulated historical state. In that case, the behavioral interpretation of current correctness would not become part of the subsequent sequence memory. CE-KT feeds the modulated hidden state and cell state into the next LSTMCell, so the state adjustment induced by current behavioral information becomes part of the later history and can influence subsequent state updating and response prediction.

4. Experimental Design

The experiments evaluate CE-KT from four complementary angles. RQ1 examines whether behavioral proxy scores remain associated with future same-skill performance after holding current response correctness fixed. RQ2 tests whether standard KT architectures show systematic prediction bias across behavioral-proxy strata. RQ3 compares correctness-specific recurrent full-state modulation with ordinary behavioral input, output-only modulation, no-gate variants, negative controls, and structural ablations. RQ4 examines whether the complete model produces differentiated predictions for interactions with the same current correctness but different behavioral contexts. All experiments use the same student-level split, training pipeline, prediction target, and evaluation rules, with explicit restrictions on item statistics, historical summaries, and future same-skill information.

4.1 Dataset, Sample Split, and Preprocessing

4.1.1 Data Source and Sample Selection

This study uses the ASSISTments 2012-2013 problem logs because they contain the information needed for KT sequence modeling: student interaction sequences, skill identifiers, response correctness labels, response time, hints, attempts, bottom-hint indicators, and affect-probability fields. After retaining tutor-mode main problems and removing records with invalid key fields, the initial valid problem-log set contained 2,574,463 records, 28,700 students, and 265 skills. Among them, 12,024 students had at least 50 valid interactions.

To obtain sufficiently long sequences and enough repeated same-skill opportunities, the experiment selected the 1,000 students with the richest interaction records and the 30 most frequent skills. Records were retained only when they satisfied both filters, resulting in 446,462 interactions. The 30 selected skills covered 62.25% of valid records before filtering, and the final sample accounted for 17.34% of the valid problem logs. This choice improves the observability of repeated same-skill opportunities and sequence length, but it also biases the sample toward high-activity students and high-frequency skills. The affect-probability fields were inherited model outputs from the dataset and were used only to assist weak-supervision anchor selection; they are not treated as self-report or human-observed affect labels.

#### 4.1.2 Core Fields and Student-Level Split

Table 1 summarizes the core fields and their experimental roles. To avoid leakage of student-specific behavioral patterns across data partitions, the data were split by student into training, validation, and test sets with 800, 100, and 100 students, respectively. All interactions from the same student belonged to only one partition. The split used a fixed random state, and all models shared the same partition.

Table 1. Core data fields and experimental use

| Field category | Experimental use |
|---|---|
| Sequence identifiers | Organize interactions by student and determine temporal order |
| Problem and skill fields | Construct item statistics, skill encodings, and next-skill prediction targets |
| Response result | Convert to a 0/1 response correctness label for base interaction input and prediction labels |
| Response process | Construct behavioral features related to response time, hints, and attempts |
| Auxiliary probabilities | Assist weak-supervision sample selection; not treated as psychological-state ground truth |

#### 4.1.3 Preprocessing, Information Separation, and Proxy Validation

Preprocessing followed the information-use rules in Section 3.2. Log response time was standardized within student using only interactions before the current time step. When the causal history was insufficient, fallback statistics estimated from the training set were used. Item-level statistics were estimated only from the training set and then frozen for validation and test students.

For RQ1, test interactions were first separated by current response correctness. Correct interactions were split into high- and low-score groups according to the median of the positive-interaction behavioral condition score, and incorrect interactions were split analogously according to the median of the negative-interaction behavioral condition score. The outcome was weighted same-skill performance in the non-overlapping fourth-to-eighth subsequent opportunities, which was not used for weak-label construction. Each side used 300 student-level bootstrap resamples, and the reported difference was calculated as the high-score group minus the low-score group, with 95% confidence intervals and empirical p-values.

### 4.2 Model Training and Common Evaluation Rules

#### 4.2.1 Training Settings

To keep structural comparisons aligned, all recurrent KT models used the same student-level split, next-interaction correctness prediction target, and training procedure. The interaction embedding formed from skill and response correctness used a dimension of 64, and the recurrent hidden-state dimension was also 64. Batch size was 64, dropout was 0.10, and the CE-KT gate amplitude was set to 0.50. Because the compared models have different architectures, they do not have identical parameter counts; parameter size is therefore reported alongside model results.

Models were trained with masked binary cross-entropy. Padded sequence positions did not contribute to the loss, and the loss was averaged only over positions with valid next-interaction prediction targets. All models used Adam with a learning rate of 0.001 and weight decay of 0.00001, and were trained for at most four epochs. The best checkpoint was selected by validation-set AUC, and the test set was not used for parameter tuning or model selection.

To account for random variation from initialization and mini-batch order, each model was trained with the same five random seeds. Means and standard deviations are reported across seeds. Stratified residual analyses and mechanism diagnostics use the average interaction-level prediction probability across the five trained models. The neural-network settings were fixed from prior development on the training and validation data rather than tuned separately for each comparison model. The influence of weak-supervision thresholds, same-skill windows, and gate amplitude is examined through sensitivity analyses.

#### 4.2.2 Metrics and Stratified Bias Analysis

Because RQ2 concerns prediction bias related to behavioral-proxy strata, evaluation cannot rely on AUC alone. The study evaluates models from three perspectives: ranking ability, probabilistic error, and calibration. AUC measures the model's ranking ability for distinguishing correct and incorrect targets. LogLoss and Brier score measure the discrepancy between predicted probabilities and observed outcomes. Expected Calibration Error (ECE) measures the agreement between predicted probabilities and observed correctness rates, using 10 equal-width probability bins. Higher AUC is better, whereas lower LogLoss, Brier score, and ECE are better.

Beyond overall metrics, each test interaction was assigned the behavioral condition score applicable to its current correctness. All interactions were sorted by this score and divided into four approximately equal strata, Q1 through Q4. For each stratum, the analysis reports the observed correctness rate, mean predicted probability, group-level residual, and mean absolute error. The group-level residual is defined as observed correctness rate minus mean predicted probability; negative values indicate overprediction, and positive values indicate underprediction.

#### 4.2.3 Uncertainty Estimation and Result Reporting

Uncertainty arises from two sources. First, neural-network training varies with initialization and mini-batch order. Second, multiple interaction records from the same student are not independent. The first source is addressed by training each model with five random seeds. The second is addressed by using students, rather than individual rows, as the resampling unit for statistical inference.

The RQ1 proxy-validation analysis does not involve repeated neural-network training and therefore uses student-level bootstrap separately within correct and incorrect interactions. For RQ2 overall model evaluation and RQ3 model comparisons, the study uses a seed-by-student two-level bootstrap. Each bootstrap draw first resamples five training seeds with replacement, then resamples test students with replacement within each selected seed, and recomputes metric differences using the same resampled student set for the complete model and the comparison model. RQ4 uses the mean interaction-level prediction across five seeds as a descriptive mechanism diagnosis and does not conduct separate significance tests for each mechanism group.

All model differences are defined as CE-KT minus the comparison model. The 95% confidence interval is computed from the 2.5th and 97.5th percentiles of the bootstrap difference distribution, and two-sided empirical p-values are reported. Raw model metrics are reported as means and standard deviations across five seeds; model comparisons report metric differences, 95% confidence intervals, and empirical p-values. With this difference convention, positive Delta AUC favors CE-KT, whereas negative Delta LogLoss, Delta Brier, and Delta ECE favor CE-KT.

### 4.3 Model Comparisons, Negative Controls, and Mechanism Tests

RQ2 uses cross-architecture comparisons to test whether behavioral-proxy stratified prediction bias is specific to DKT or can also be observed in different sequence-modeling mechanisms. RQ3 compares the complete model with alternative behavior-fusion approaches and uses negative controls, structural ablations, and gate-input ablations to analyze the dependence of predictive performance on recurrent feedback, gate structure, and gate-conditioning information. RQ4 uses mechanism-group diagnostics to examine whether the complete model treats interactions with the same current correctness differently according to behavioral context. Unless otherwise stated, all models share the same student split, random seeds, prediction target, and evaluation rules.

### 4.3.1 Main Comparison Models and Validation Purpose

The comparison models were selected for architectural representativeness, input comparability, and validation purpose, rather than as an exhaustive benchmark of all KT methods. For RQ2, DKT, DKVMN, SAKT, and AKT-S represent recurrent-state, external-memory, self-attention, and monotonic-distance attention mechanisms, respectively. They are compared with CE-KT under the same stratified residual analysis. All models use the same skill-level input. SAKT and AKT-S use the most recent 200 interactions as the history window. AKT-S is adapted to the current skill-level data and does not include the original AKT item-difficulty component, so it is used only as a cross-architecture diagnostic baseline. DKVMN, SAKT, and AKT-S are not used to attribute the effect of the proposed gate mechanism.

For RQ3, the main comparisons are more directly mechanism-matched. DKT serves as the classical baseline. DKT+FullBehavior tests whether directly concatenating the full behavioral feature vector to the sequence input is sufficient. CE-KT-NoGate uses the same stepwise LSTMCell update implementation as the complete model but removes the gate, ruling out the explanation that performance changes are caused merely by replacing a batched LSTM with LSTMCell. CE-KT-OutputOnly modulates the current prediction only and does not feed the modulated state back into subsequent time steps, testing the role of recurrent feedback.

### 4.3.2 Negative Controls

The main model comparison alone cannot rule out explanations based on parameter scale or the marginal distribution of behavioral information. Two negative controls are therefore included: CE-KT-PermutedContext and CE-KT-RandomProxy. CE-KT-PermutedContext permutes gate-conditioning information among interactions from the same student and the same current correctness group. This preserves the model structure and the marginal distribution of behavioral information while breaking the pairing between an interaction and its own behavioral context. It tests whether the model depends on the true interaction-level behavioral context. CE-KT-RandomProxy keeps the full behavioral features and separate correct/incorrect gates, but replaces the three behavioral proxy scores with random values. It tests whether the proxy scores provide stable predictive increment beyond the full behavioral feature vector.

### 4.3.3 Structural Ablations and Their Validation Purpose

Structural ablations examine two aspects: correctness-specific gate paths and the combination of multiplicative scaling and additive shifting. CE-KT-SharedGate uses a single gate network for both correct and incorrect interactions, testing whether separate routing is needed. CE-KT-PosOnly enables gating only for correct interactions, and CE-KT-NegOnly enables gating only for incorrect interactions, testing whether one-sided modulation is sufficient. CE-KT-GammaOnly removes the additive shift and retains only multiplicative scaling, whereas CE-KT-BetaOnly removes multiplicative scaling and retains only additive shifting. These variants test the separate contributions of the two modulation operations.

All structural ablations retain recurrent feedback; the modulated state is passed to the next time step. Because SharedGate, PosOnly, and NegOnly reduce gate paths and also reduce parameter counts, comparisons with the complete model should be interpreted as structural-deletion comparisons rather than strictly parameter-matched causal attributions. GammaOnly and BetaOnly use symmetric dual-gate deletion structures and are more directly targeted to the two state-modulation operations, but even these comparisons are not presented as definitive causal isolation of one operation.

#### 4.3.4 Gate-Input Source Ablation

To distinguish the role of the three behavioral condition scores from the role of full behavioral features, the study includes gate-input source ablations. CE-KT-ProxyOnly uses only the low-trust behavioral similarity score, positive-interaction behavioral condition score, and negative-interaction behavioral condition score as gate conditions. CE-KT-FullContext retains the 13-dimensional full behavioral feature vector in addition to the three proxy scores. The two variants share the same recurrent dual-gate structure, hidden dimension, number of training epochs, student split, random seeds, and prediction target; the main difference is whether the gate network directly reads the full behavioral feature vector.

DKT+FullBehavior is included as the ordinary behavioral-input baseline, allowing the analysis to test whether proxy-conditioned recurrent gating improves prediction relative to feature concatenation. Gate-input source ablations are evaluated by AUC, LogLoss, Brier score, and ECE, and model differences are estimated using the seed-by-student two-level bootstrap. The analysis also reports model parameter counts, failed seeds, and any stabilization handling. Q1-Q4 residuals are used only as supplementary stratified diagnostics, not as the sole criterion for deciding whether proxy information is sufficient.

#### 4.3.5 Positive- and Negative-Side Mechanism Diagnostics

RQ4 examines whether the model differentiates interactions with the same current response correctness according to behavioral context. Mechanism groups are constructed separately within currently correct and currently incorrect interactions. For incorrect interactions, the 25th and 75th percentiles of the negative-interaction behavioral condition score define the low-score error group and the post-error improvement candidate group. For correct interactions, the analogous percentiles define low- and high-score correct-interaction groups. Interactions between the 25th and 75th percentiles are not included in this mechanism diagnosis.

These group names are operational labels based on behavioral proxy scores and weak-supervision anchors. They do not mean that the corresponding interactions have been externally validated as low-quality correctness, true mastery, or post-error improvement cases. RQ4 is a descriptive mechanism diagnosis and is not used to confirm true cognitive types or causal effects. For each group, the analysis reports the observed target correctness rate, mean model prediction, residuals, CE-KT minus DKT movement, and absolute-bias reduction to assess whether prediction shifts move toward the observed group rate. This analysis remains descriptive and does not confirm true cognitive types or causal effects.

## 5. Results

### 5.1 RQ1: Behavioral Heterogeneity Within Fixed Correctness

RQ1 examined whether behavioral condition scores were associated with future same-skill performance after interactions were separated by current response correctness. As shown in Table 2, among currently correct interactions, the high-score group had a student-bootstrap mean difference of -0.0111 relative to the low-score group, with a 95% confidence interval of [-0.0207, -0.0005] and $p = .038$. Although the confidence interval did not include zero, the magnitude was only about 1.11 percentage points and the direction was negative. Therefore, this result indicates a small behavioral association within correct responses, but it does not support interpreting higher positive-interaction scores as stronger evidence of stable mastery.

Among currently incorrect interactions, the high-score group exceeded the low-score group by 0.0907 in future same-skill performance, with a 95% confidence interval of [0.0151, 0.1995] and $p = .006$. This result supports an association between the negative-interaction behavioral condition score and later same-skill performance differences within incorrect responses. Overall, RQ1 supports behavioral heterogeneity within fixed response correctness, but the positive and negative sides differ in direction, magnitude, and interpretability.

Table 2. Independent outcome validity of behavioral condition scores for correct and incorrect interactions

| Condition score | Current correctness | Raw low group | Raw high group | Bootstrap high-low | 95% CI | p |
|---|---|---|---|---|---|---|
| Positive-interaction behavioral condition score | 1 | 0.7712 | 0.7601 | -0.0111 | [-0.0207, -0.0005] | 0.038 |
| Negative-interaction behavioral condition score | 0 | 0.5472 | 0.6478 | +0.0907 | [0.0151, 0.1995] | 0.006 |

**5.2 RQ2: Stratified Prediction Bias Across Behavioral Proxy Scores**

RQ2 tested whether standard KT models with different sequence-modeling mechanisms showed systematic prediction bias across behavioral-proxy strata. Because the positive- and negative-side scores were trained from different weak-supervision tasks, their values are not directly comparable. The mixed Q1-Q4 strata are therefore used only to examine whether prediction residuals vary across behavioral-proxy regions; they do not represent a unified response-quality scale.

For overall probabilistic metrics, CE-KT obtained an AUC of 0.7144, LogLoss of 0.5376, Brier score of 0.1788, and ECE of 0.0144. Relative to DKT, AUC increased by 0.0294, while LogLoss and Brier score decreased by 0.0168 and 0.0070, respectively. The corresponding two-level bootstrap confidence intervals were [0.0252, 0.0340], [-0.0209, -0.0135], and [-0.0088, -0.0055], none of which included zero. ECE decreased by 0.0032, but its confidence interval, [-0.0082, 0.0019], included zero. Thus, CE-KT reliably improved ranking ability and reduced probabilistic error relative to DKT, but did not show a stable overall calibration advantage in this comparison.

Table 3. Conditional-proxy quartile results for DKT and CE-KT

| Stratum | Observed rate | DKT pred. | DKT residual | CE-KT pred. | CE-KT residual |
|---|---|---|---|---|---|
| Q1 | 0.6464 | 0.6933 | -0.0469 | 0.6846 | -0.0381 |
| Q2 | 0.7140 | 0.7109 | 0.0031 | 0.7290 | -0.0150 |
| Q3 | 0.7268 | 0.7089 | 0.0180 | 0.7151 | 0.0117 |
| Q4 | 0.7651 | 0.7557 | 0.0094 | 0.7646 | 0.0005 |

The descriptive residuals in Table 3 show that DKT errors were not evenly distributed across proxy-score strata. In Q1, the observed correctness rate was 0.6464, whereas DKT predicted 0.6933, corresponding to overprediction by 4.69 percentage points. In Q3 and Q4, DKT underpredicted by 1.80 and 0.94 percentage points, respectively. CE-KT reduced the Q1 overprediction from 4.69 to 3.81 percentage points and reduced the Q3 and Q4 underprediction to 1.17 and 0.05 percentage points. CE-KT also lowered the sample-level mean absolute error across the four strata. However, Q2 changed from a residual of 0.0031 to -0.0150, so not all strata improved. Because the strata mix two different weak-supervision tasks, these residuals should not be interpreted as a monotonic scale of mastery or response quality. To examine whether the stratified bias was specific to DKT or appeared across architectures, the next analysis compares DKT, DKVMN, SAKT, AKT-S, and CE-KT under the same Q1-Q4 stratification.

Table 4. Conditional-proxy quartile residuals across KT architectures

| Model | Q1 residual | Q2 residual | Q3 residual | Q4 residual | Max gap | Clustered Wald p |
|---|---|---|---|---|---|---|
| DKT | -0.0469 | 0.0031 | 0.0180 | 0.0094 | 0.0649 | 0.00136 |
| DKVMN | -0.0439 | -0.0037 | 0.0244 | 0.0206 | 0.0683 | <0.000001 |
| SAKT | -0.0569 | -0.0043 | 0.0101 | 0.0190 | 0.0760 | <0.000001 |
| AKT-S | -0.0331 | -0.0112 | 0.0177 | 0.0121 | 0.0509 | 0.00322 |

| CE-KT (FullGate) | -0.0381 | -0.0150 | 0.0117 | 0.0005 | 0.0498 | 0.000823 |
|---|---|---|---|---|---|---|

The cross-architecture test in Table 4 used a joint Wald test of residuals on Q1-Q4 indicators with student-clustered robust covariance. DKT, DKVMN, SAKT, and AKT-S all showed statistically reliable residual differences across strata, with maximum cross-stratum residual gaps of 0.0649, 0.0683, 0.0760, and 0.0509, respectively. Across the four standard KT architectures, Q1 showed overprediction, whereas Q3 or Q4 showed underprediction to varying degrees. These results indicate that stratified residual bias was not unique to DKT; it also appeared in external-memory and attention-based KT models, although the magnitude and stratum pattern differed across architectures.

CE-KT had the smallest maximum cross-stratum residual gap among the five models, at 0.0498. However, its cross-stratum Wald test remained significant ($p = .000823$), indicating that stratified residual differences were reduced but not eliminated. SAKT and AKT-S used a 200-step attention window. AKT-S was adapted to the 30-skill input setting and retained causal monotonic distance decay, but did not include the original AKT item-difficulty component; it should therefore be interpreted as a diagnostic cross-architecture baseline rather than a full reproduction of AKT.

### 5.3 RQ3: Recurrent Correctness-Specific State Modulation Versus Other Behavior-Fusion Strategies

RQ3 compared correctness-specific recurrent state modulation with alternatives that use behavior as ordinary input or modulate only the current output. The main comparisons examined overall differences, negative controls tested alternative explanations, structural ablations tested gate separation and scale-shift modulation, and gate-input ablations examined the information sources used by the gate.

#### 5.3.1 Main Model Comparisons

Within the mechanism-matched RQ3 comparisons in Table 5, CE-KT outperformed DKT, DKT+FullBehavior, CE-KT-OutputOnly, and CE-KT-NoGate on AUC, LogLoss, and Brier score. Its mean ECE was also lower than these mechanism-matched baselines, but the corresponding bootstrap confidence intervals included zero, so the complete model cannot be claimed to provide a stable overall calibration advantage. DKVMN, SAKT, and AKT-S are treated as cross-architecture performance context and are not used for gate-mechanism attribution.

Table 5. Five-seed results for main comparison models

| **Model** | **AUC** | **LogLoss** | **Brier** | **ECE** |
|---|---|---|---|---|
| DKVMN | 0.7156 ± 0.0025 | 0.5337 ± 0.0020 | 0.1774 ± 0.0008 | 0.0088 ± 0.0034 |
| DKT | 0.6845 ± 0.0047 | 0.5546 ± 0.0023 | 0.1858 ± 0.0010 | 0.0163 ± 0.0033 |
| DKT+FullBehavior | 0.6936 ± 0.0012 | 0.5504 ± 0.0006 | 0.1841 ± 0.0003 | 0.0179 ± 0.0040 |
| CE-KT-OutputOnly | 0.6932 ± 0.0050 | 0.5493 ± 0.0025 | 0.1835 ± 0.0010 | 0.0147 ± 0.0036 |
| CE-KT-NoGate | 0.6844 ± 0.0047 | 0.5546 ± 0.0023 | 0.1859 ± 0.0010 | 0.0162 ± 0.0032 |
| CE-KT-PermutedContext | 0.7115 ± 0.0016 | 0.5396 ± 0.0009 | 0.1797 ± 0.0003 | 0.0096 ± 0.0016 |
| CE-KT (FullGate) | 0.7144 ± 0.0019 | 0.5376 ± 0.0010 | 0.1788 ± 0.0003 | 0.0144 ± 0.0040 |
| SAKT | 0.6981 ± 0.0011 | 0.5456 ± 0.0012 | 0.1823 ± 0.0006 | 0.0109 ± 0.0055 |
| AKT-S | 0.6894 ± 0.0005 | 0.5498 ± 0.0004 | 0.1837 ± 0.0002 | 0.0103 ± 0.0030 |

The two-level bootstrap results in Table 6 show reliable AUC, LogLoss, and Brier improvements over DKT, DKT+FullBehavior, CE-KT-OutputOnly, and CE-KT-NoGate. Relative to DKT, CE-KT improved AUC by 0.0294 and reduced LogLoss and Brier score by 0.0168 and 0.0070. Relative to DKT+FullBehavior, AUC improved by 0.0206, while LogLoss and Brier score decreased by 0.0128 and 0.0053. Relative to CE-KT-OutputOnly, AUC improved by 0.0208, LogLoss decreased by 0.0116, and Brier score decreased by 0.0046. Relative to CE-KT-NoGate, AUC improved by 0.0295, LogLoss decreased by 0.0169, and Brier score decreased by 0.0070. These comparisons indicate that the performance gain was not only due to adding behavioral input; recurrent feedback and the complete gate structure were associated with better ranking and probabilistic error. However, ECE confidence intervals included zero in these four comparisons, so stable calibration improvement cannot be concluded.

As cross-architecture context, CE-KT reliably improved AUC, LogLoss, and Brier score relative to SAKT and AKT-S, while ECE differences included zero. Relative to DKVMN, the AUC difference was -0.0010 with a 95% confidence interval of [-0.0037, 0.0014], indicating no reliable AUC difference. LogLoss, Brier score, and ECE differences relative to DKVMN were 0.0038, 0.0014, and 0.0056, respectively, and their confidence intervals were entirely above zero. Because lower values are better for these metrics, CE-KT was reliably worse than DKVMN on LogLoss, Brier score, and ECE. Therefore, CE-KT should not be described as the overall strongest model.

Table 6. Two-level bootstrap differences for CE-KT relative to main comparisons

| **Baseline** | **Delta AUC [95% CI]** | **Delta LogLoss [95% CI]** | **Delta Brier [95% CI]** | **Delta ECE [95% CI]** |
|---|---|---|---|---|
| DKT | 0.0294 [0.0252, 0.0340] | -0.0168 [-0.0209, -0.0135] | -0.0070 [-0.0088, -0.0055] | -0.0032 [-0.0082, 0.0019] |
| DKT+FullBehavior | 0.0206 [0.0179, 0.0232] | -0.0128 [-0.0152, -0.0107] | -0.0053 [-0.0063, -0.0044] | -0.0043 [-0.0104, 0.0027] |
| CE-KT-OutputOnly | 0.0208 [0.0160, 0.0255] | -0.0116 [-0.0146, -0.0087] | -0.0046 [-0.0059, -0.0035] | -0.0018 [-0.0050, 0.0014] |
| CE-KT-NoGate | 0.0295 [0.0253, 0.0342] | -0.0169 [-0.0210, -0.0135] | -0.0070 [-0.0089, -0.0055] | -0.0030 [-0.0079, 0.0019] |
| CE-KT-PermutedContext | 0.0029 [0.0019, 0.0038] | -0.0020 [-0.0026, -0.0013] | -0.0009 [-0.0012, -0.0006] | 0.0031 [0.0000, 0.0067] |
| SAKT | 0.0167 [0.0128, 0.0205] | -0.0080 [-0.0107, -0.0048] | -0.0035 [-0.0047, -0.0021] | 0.0031 [-0.0034, 0.0099] |
| AKT-S | 0.0247 [0.0223, 0.0274] | -0.0121 [-0.0137, -0.0106] | -0.0049 [-0.0055, -0.0043] | 0.0020 [-0.0040, 0.0072] |
| DKVMN | -0.0010 [-0.0037, 0.0014] | 0.0038 [0.0012, 0.0075] | 0.0014 [0.0003, 0.0030] | 0.0056 [0.0008, 0.0117] |

### 5.3.2 Negative Control Results

Relative to CE-KT-PermutedContext, CE-KT improved AUC by only 0.0012, with a 95% confidence interval of [0.0005, 0.0021], and LogLoss and Brier score also showed small but reliable improvements. The ECE difference included zero. Because CE-KT-PermutedContext still achieved an AUC of 0.7115 after breaking the pairing between each interaction and its own behavioral context, the true interaction-level pairing appears to provide only a small measurable contribution. The overall gain cannot be attributed entirely to precise behavioral-context pairing.

CE-KT-RandomProxy retained the full behavioral features and model structure but replaced the three behavioral proxy scores with random values. Its AUC was 0.7120. The complete model improved AUC over CE-KT-RandomProxy by only 0.0006, with a 95% confidence interval of [-0.0005, 0.0016], and the LogLoss, Brier, and ECE differences also included zero. Thus, when the full behavioral feature vector is already available to the gate network, the current results do not show that the three proxy scores provide a stable additional predictive increment. This motivates the ProxyOnly ablation, which tests whether the proxy scores can support recurrent state modulation when full behavioral features are removed.

### 5.3.3 Structural Ablation Results

The structural ablations in Table 7 further test whether separate correct/incorrect gate paths, two-sided modulation, and the combination of multiplicative scaling and additive shifting contribute to the complete model.

Table 7. Negative control and structural ablation results across five seeds

| Model | AUC | LogLoss | Brier | ECE |
|---|---|---|---|---|
| CE-KT-RandomProxy | 0.7120 ± 0.0019 | 0.5388 ± 0.0010 | 0.1793 ± 0.0004 | 0.0116 ± 0.0014 |
| CE-KT-SharedGate | 0.6993 ± 0.0012 | 0.5466 ± 0.0008 | 0.1825 ± 0.0003 | 0.0142 ± 0.0014 |
| CE-KT-PosOnly | 0.7006 ± 0.0009 | 0.5470 ± 0.0004 | 0.1827 ± 0.0002 | 0.0189 ± 0.0035 |
| CE-KT-NegOnly | 0.6997 ± 0.0042 | 0.5451 ± 0.0020 | 0.1817 ± 0.0008 | 0.0136 ± 0.0039 |
| CE-KT-GammaOnly | 0.7016 ± 0.0041 | 0.5464 ± 0.0022 | 0.1825 ± 0.0008 | 0.0175 ± 0.0018 |
| CE-KT-BetaOnly | 0.7084 ± 0.0013 | 0.5404 ± 0.0006 | 0.1799 ± 0.0002 | 0.0100 ± 0.0020 |
| CE-KT (FullGate) | 0.7127 ± 0.0016 | 0.5384 ± 0.0006 | 0.1792 ± 0.0002 | 0.0101 ± 0.0031 |

CE-KT improved AUC over CE-KT-SharedGate, CE-KT-PosOnly, and CE-KT-NegOnly by 0.0133, 0.0120, and 0.0129, respectively, and LogLoss and Brier score were also reliably lower. These results support the complete dual-gate configuration relative to corresponding deletion variants. However, because these ablations have fewer parameters than the complete model, the differences cannot be attributed strictly to correctness-specific routing alone.

For modulation-operation ablations, CE-KT improved AUC by 0.0110 over CE-KT-GammaOnly and by 0.0042 over CE-KT-BetaOnly, with confidence intervals excluding zero. LogLoss and Brier score also improved. For ECE, CE-KT improved over GammaOnly but did not show a reliable difference relative to BetaOnly. Thus, the scale-plus-shift combination mainly improved ranking and probabilistic error, while calibration gains were not stable across all ablation comparisons.

In summary, recurrent feedback and complete state modulation outperformed ordinary behavioral input, CE-KT-OutputOnly, and CE-KT-NoGate. At the same time, CE-KT-RandomProxy was not reliably different from CE-KT, and SharedGate and one-sided gates differed in parameter count. The current evidence therefore supports the full recurrent gate configuration as a strong predictive structure over mechanism-matched baselines, but it does not uniquely attribute the improvements to the three proxy scores, correctness-specific routing, or any single modulation operation.

### 5.3.4 Gate-Input Source Ablation Results

Although the above results support the overall predictive advantage of the complete recurrent gate configuration, they do not specify which behavioral information sources are necessary. Table 8 compares DKT+FullBehavior, CE-KT-ProxyOnly, and CE-KT-FullContext to test whether the three proxy scores alone can support recurrent gating and whether adding full behavioral features provides further improvement.

Table 8. Gate-input source ablation results across five seeds

| Model | AUC | LogLoss | Brier | ECE |
|---|---|---|---|---|
| DKT+FullBehavior | 0.6936 ± 0.0012 | 0.5504 ± 0.0006 | 0.1841 ± 0.0003 | 0.0179 ± 0.0040 |
| CE-KT-ProxyOnly | 0.7077 ± 0.0025 | 0.5419 ± 0.0016 | 0.1806 ± 0.0007 | 0.0139 ± 0.0027 |
| CE-KT-FullContext | 0.7113 ± 0.0016 | 0.5393 ± 0.0009 | 0.1795 ± 0.0004 | 0.0121 ± 0.0025 |

CE-KT-ProxyOnly achieved an AUC of 0.7077, higher than DKT+FullBehavior at 0.6936. LogLoss and Brier score also decreased from 0.5504 and 0.1841 to 0.5419 and 0.1806. The two-level bootstrap showed an AUC difference of 0.0140 with a 95% confidence interval of [0.0114, 0.0166]. LogLoss and Brier differences were -0.0086 and -0.0035, with 95% confidence intervals of [-0.0105, -0.0070] and [-0.0043, -0.0028], respectively. These results show that the three proxy scores alone can support recurrent gating that outperforms ordinary behavioral feature concatenation on ranking and probabilistic error. However, the ECE confidence interval included zero, so stable calibration improvement is not established.

Within the same gate structure, CE-KT-FullContext further improved AUC over CE-KT-ProxyOnly by 0.0035 and reduced LogLoss and Brier score by 0.0027 and 0.0012. The 95% confidence intervals for these three differences excluded zero. ECE decreased by 0.0015, but its confidence interval included zero. Thus, the proxy scores contain useful compressed behavioral information, but they do not fully summarize the predictive information in the original behavioral features. Response time, hint use, attempt count, and historical behavior still provide additional ranking and probabilistic-error gains beyond the proxy scores.

### 5.4 RQ4: Mechanism Diagnosis Within Correct and Incorrect Interactions

RQ4 provides a descriptive mechanism diagnosis of CE-KT's gate behavior. It does not claim that the model identifies true cognitive types. The analysis determines whether average predicted probabilities move in expected directions across four operational groups and whether each model's predictions are closer to the observed target correctness rate. If CE-KT moves predictions closer to the observed group rate, the gate modulation is treated as consistent with the expected correction for that group; otherwise, the corresponding proxy signal has not been stably converted into a reasonable prediction.

Table 9. Observed target rates and mean model predictions for four mechanism groups

| **Mechanism group** | **Observed rate** | **DKT** | **DKVMN** | **SAKT** | **AKT-S** | **CE-KT (FullGate)** |
|---|---|---|---|---|---|---|
| Low-score correct interactions | 0.7785 | 0.7588 | 0.7779 | 0.7635 | 0.7807 | 0.7948 |
| High-score correct interactions | 0.7916 | 0.7834 | 0.7721 | 0.7624 | 0.7843 | 0.7981 |
| Low-score error interactions | 0.4946 | 0.6106 | 0.5998 | 0.6354 | 0.5566 | 0.5533 |
| Post-error improvement candidate group | 0.6278 | 0.6171 | 0.6038 | 0.6676 | 0.5917 | 0.5903 |

Table 10. Model residuals for four mechanism groups (observed rate minus predicted probability)

| **Mechanism group** | **DKT** | **DKVMN** | **SAKT** | **AKT-S** | **CE-KT (FullGate)** |
|---|---|---|---|---|---|
| Low-score correct interactions | 0.0197 | 0.0006 | 0.0150 | -0.0021 | -0.0163 |
| High-score correct interactions | 0.0082 | 0.0195 | 0.0292 | 0.0073 | -0.0065 |
| Low-score error interactions | -0.1159 | -0.1052 | -0.1407 | -0.0619 | -0.0587 |
| Post-error improvement candidate group | 0.0107 | 0.0240 | -0.0398 | 0.0362 | 0.0375 |

For correct interactions, the low- and high-score groups had observed target rates of 0.7785 and 0.7916, a difference of about 1.31 percentage points. Given the small and directionally complex RQ1 result for the positive side, this difference should be interpreted only as a descriptive next-target prediction pattern. CE-KT increased predictions for both groups and reduced absolute bias by 0.34 and 0.17 percentage points, respectively. This indicates group-specific modulation, but the magnitude is too small to show that the model has stably identified evidential-strength differences within correct interactions.

For incorrect interactions, CE-KT showed its clearest correction in the low-score error group. The observed target rate was 0.4946, whereas DKT predicted 0.6106, overpredicting by 11.59 percentage points.

CE-KT reduced the prediction to 0.5533, lowering the overprediction to 5.87 percentage points and reducing absolute bias by 5.73 percentage points. This result indicates that the incorrect-response gate helped suppress the baseline model's overoptimistic prediction for this operational low-score error group. The group label remains proxy-based and does not identify an externally validated cognitive type.

The post-error improvement candidate group had an observed target rate of 0.6278, higher than the low-score error group. However, CE-KT did not correct this group in the expected direction. Compared with DKT's prediction of 0.6171, CE-KT decreased the prediction to 0.5903, increasing absolute bias from 1.07 to 3.75 percentage points. Thus, the current incorrect-side gate captured the overoptimism problem in low-score errors more clearly than it captured the higher subsequent performance signal in post-error improvement candidates. In this group, DKT had the smallest absolute residual.

Across the four mechanism groups, the best model differed by group: DKVMN had the smallest absolute residual in the low-score correct group, CE-KT had the smallest absolute residual in the high-score correct and low-score error groups, and DKT had the smallest absolute residual in the post-error improvement candidate group. RQ4 therefore provides partial support and boundary evidence for the gate mechanism. CE-KT can generate behavior-conditioned group-level modulation and improves the low-score error group, but positive-side modulation is small and post-error improvement candidates are not corrected as expected.

### 5.5 Robustness and Sensitivity Analyses

To examine whether the main conclusions depend on a single operational threshold, the study varied the rapid-guessing percentile, the post-error improvement threshold, and the gate amplitude, retraining the complete model with the same data split and five random seeds. In Table 11, the negative-side future-performance difference continues to use the non-overlapping fourth-to-eighth subsequent same-skill opportunities.

Table 11. Sensitivity results for key thresholds and gate amplitude

| Setting | AUC | LogLoss | Brier | ECE | Negative-side independent difference [95% CI] |
|---|---|---|---|---|---|
| Default | 0.7127 | 0.5384 | 0.1792 | 0.0101 | 0.0907 [0.0151, 0.1995] |
| Rapid-guess percentile = 0.01 | 0.7127 | 0.5385 | 0.1792 | 0.0103 | 0.0948 [0.0145, 0.2161] |
| Rapid-guess percentile = 0.10 | 0.7125 | 0.5385 | 0.1792 | 0.0105 | 0.0900 [0.0095, 0.2091] |
| Post-error improvement threshold = 0.10 | 0.7127 | 0.5384 | 0.1792 | 0.0099 | 0.0881 [0.0088, 0.2077] |
| Post-error improvement threshold = 0.30 | 0.7127 | 0.5384 | 0.1792 | 0.0100 | 0.0872 [0.0055, 0.2084] |
| Gate amplitude = 0.25 | 0.7080 | 0.5416 | 0.1804 | 0.0129 | 0.0913 [0.0101, 0.2134] |
| Gate amplitude = 0.75 | 0.7144 | 0.5376 | 0.1788 | 0.0144 | 0.0913 [0.0101, 0.2134] |

When the rapid-guessing percentile changed from the default 5% to 1% or 10%, bootstrap confidence intervals for AUC, LogLoss, Brier score, and ECE relative to the default setting included zero. The negative-side future-performance differences remained positive at 0.0948 and 0.0900. When the post-error improvement threshold changed from 0.20 to 0.10 or 0.30, the four model metrics again showed no stable changes, and the negative-side future-performance difference remained between 0.0872 and 0.0881 with confidence intervals above zero. These results indicate that the main model performance and the negative-side validation result do not depend on a single rapid-guessing or post-error improvement threshold.

Gate amplitude had a more visible influence. An amplitude of 0.25 reduced AUC to 0.7080 and increased LogLoss, Brier score, and ECE, suggesting that overly weak modulation was insufficient for the recurrent gate to operate effectively. An amplitude of 0.75 increased AUC to 0.7144 and reduced Brier score, but ECE increased to 0.0144, indicating a trade-off between ranking and calibration. The default amplitude of 0.50 should therefore be understood as a compromise between probabilistic error and calibration rather than as a uniquely optimal value.

Additional proxy-construction diagnostics compared same-skill windows of 2, 3, and 5 and decay weights of 0.5, 0.6, and 0.8. Correlations with the default local-gain definition ranged from 0.9319 to 1.0000, and post-error improvement candidates accounted for 21.41% to 28.44% of incorrect interactions. These results suggest that proxy rankings were generally consistent, although candidate coverage changed with the window and decay settings. Overall, changing the rapid-guessing and post-error improvement thresholds did not alter the main conclusions, and the negative-side independent validity remained stable. Gate amplitude, however, showed an interpretable trade-off between performance and calibration.

## 6. Discussion and Conclusion

The analyses show that, after holding response correctness fixed, incorrect interactions contained long-range performance differences that were associated with behavioral condition scores and aligned with the expected direction. Correct interactions also showed a small behavior-related difference, but the direction was negative and therefore did not support interpreting the score as evidence of stable mastery. Overall, CE-KT outperformed ordinary behavioral-input models, output-only modulation, NoGate, and the SAKT performance context, and it reduced bias in most conditional-score regions. However, the learned behavioral proxy scores did not show a stable independent increment over CE-KT-RandomProxy, and calibration worsened for the post-error improvement candidate group. Therefore, the main contribution of this study is to propose and test a functionally explicit recurrent behavioral-conditioning pathway, not to claim that the model has measured true cognitive states or fully implemented the intended mechanism.

### 6.1 Within-Correctness Differences and Their Meaning

RQ1 provides clearer support on the incorrect side: the high-score group for incorrect interactions showed higher performance in the non-overlapping future window. On the correct side, the statistically distinguishable difference was small and negative, which does not support the interpretation that high-score correct interactions indicate more stable mastery. Thus, RQ1 supports behavior-related heterogeneity within response correctness, but the semantics of the correct-side proxy still need to be reconstructed. It should not be interpreted as a true measure of mastery, guessing, or learning gain.

### 6.2 Conditional Stratification Bias in KT Models

The mixed Q1-Q4 analysis showed that the direction of DKT prediction error varied across conditional-proxy regions. CE-KT improved most regions and the overall probabilistic metrics, but it did not improve Q2. Because the two sides of the score were derived from different weak-supervision tasks, the current result should be described as descriptive conditional stratification rather than as a unified observation-quality scale. When evaluating behavior-enhanced KT models, ranking metrics, probabilistic error, calibration, and grouped residuals should be reported together.

### 6.3 Structural Role of the Recurrent Gate and Attribution Boundaries

RQ3 indicates that placing behavioral information in recurrent state modulation generally outperformed ordinary behavioral input and output-only modulation. CE-KT-RandomProxy did not show that the three proxy scores had a stable marginal increment beyond the full behavioral feature package. However, CE-KT-ProxyOnly remained significantly better than DKT+FullBehavior after the full behavioral inputs were removed, indicating that the three proxy scores contain compressed information that can be used for recurrent modulation. At the same time, CE-KT remained significantly better than CE-KT-ProxyOnly on AUC, LogLoss, and Brier score, showing that the proxy scores did not fully preserve the original behavioral evidence. Therefore, the evidence supports the conclusion that proxy compression is useful but insufficient; it does not support the stronger claims that the model gain comes entirely from the three proxies or that the original behavioral features can be fully removed.

### 6.4 Effective and Ineffective Parts of the Mechanism

CE-KT reduced the overprediction for the low-score error group, but it further lowered predictions for the post-error improvement candidate group and did not preserve that group's higher observed target rate. Improvements on the correct side were also small. This pattern indicates that the network more readily learned stronger negative modulation for some errors, but it had not yet learned to preserve the higher subsequent-performance candidates within incorrect interactions.

### 6.5 Contributions, Limitations, and Future Work

This study treats behavioral process information as conditioning information for response correctness, proposes a recurrent full-state gate with separate correct- and incorrect-response pathways, and constrains attribution through multiple metrics, student- and seed-aware bootstrap comparisons, negative controls, ablations, sensitivity analyses, and mechanism-group diagnostics. The limitations are also direct. The experiments used only high-activity students and frequent skills from ASSISTments; the problem log fields did not reconstruct the full action order; affect probabilities were not revalidated on the current analytic sample; the proxies depended on heuristic anchors; proxy scores for training students were not estimated with student-level cross-fitting; RQ1 did not jointly adjust for student, skill, item, and opportunity count; the correct-side proxy lacked independent validity; the CE-KT-RandomProxy difference was unstable; the mixed Q1-Q4 strata were not a unified quality scale; and observational log data cannot support causal inference. Future work should prioritize these checks and replicate the findings across more datasets.